\documentclass[%
 reprint,
superscriptaddress,
showpacs,preprintnumbers,
 amsmath,amssymb,
 aps,
prb,
]{revtex4-1}

\usepackage{graphicx}% Include figure files
\usepackage{dcolumn}% Align table columns on decimal point
\usepackage{bm}% bold math
\usepackage[mathlines]{lineno}% Enable numbering of text and display math
\begin{document}

\preprint{APS/123-QED}

\title{Enhanced circular photogalvanic effect in HgTe quantum wells in the heavily inverted regime}% Force line breaks with \\

\author{Jun Li}
\email{lijun@xmu.edu.cn}
\affiliation{
 Department of Physics, Semiconductor Photonics Research Center, Xiamen University, Xiamen 361005, China
}

\author{Wen Yang}
\affiliation{Beijing Computational Science Research Center, Beijing 100089, China
}
\author{Jiang-Tao Liu}
\affiliation{Department of Physics, Nanchang University, Nanchang 330031, China
}
\affiliation{Nanoscale Science and Technology  Laboratory, Institute for Advanced Study, Nanchang University, Nanchang 330031, China
}
\author{Wei Huang}

\author{Cheng Li}

\author{Song-Yan Chen}
\affiliation{
 Department of Physics, Semiconductor Photonics Research Center, Xiamen University, Xiamen 361005, China
}
\date{\today}% It is always \today, today,
\begin{abstract}
HgTe-based quantum wells (QWs) possess very strong spin-orbit interaction (SOI) and have become an ideal platform for the study of fundamental SOI-dependent phenomena and the topological insulator phase.
Circular photogalvanic effect (CPGE) in HgTe QWs is of great interest because it provides an effective optical access to probe the spin-related information of HgTe systems. However, the complex band structure and large spin-splitting of HgTe QWs makes it inadequate to analyze the experimental results of CPGE in HgTe QWs [B. Wittmann \textit{et al.}, Semicond. Sci. Technol. 25, 095005 (2010)] with reduced band models. Here, based on the realistic eight-band $\mathbf{k}\cdot\mathbf{p}$ Hamiltonian and combined with the density-matrix formalism, we present a detailed theoretical investigation of CPGE in (001)-oriented Hg$_{0.3}$Cd$_{0.7}$Te/HgTe/Hg$_{0.3}$Cd$_{0.7}$Te QWs.
We find the CPGE currents in HgTe QWs in the heavily inverted regime are significantly enhanced due to the strong distortion of band dispersion at a certain range of the energy spectrum. This enhancement effect could offer an experimental certificate that the HgTe QW is in the heavily inverted phase (usually accompanied with the emergence of two-dimensional topological edge states), and could also be utilized in engineering the high efficiency ellipticity detector of infrared and terahertz radiation [S. N. Danilov et al., J. Appl. Phys. 105, 013106 (2009)]. Additionally, within the same theoretical framework, we also investigate the interplay effect of structure inversion asymmetry and bulk inversion asymmetry and the pure spin currents driven by linearly polarized light in HgTe QWs.
\end{abstract}

\pacs{78.20.Bh, 72.25.Fe, 78.67.De, 71.28.+d, 73.21.Fg}% PACS, the Physics and Astronomy

\maketitle

\section{Introduction \label{sec:Introcution}}

HgTe, CdTe and their alloy Hg$_{1-x}$Cd$_{x}$Te($x\in[0,1]$) can comprise heterostructures with a tunable direct band-gap spanning shortwave infrared to terahertz region, and have been widely used in the devices of infrared photodetection \cite{Rogalski_MCTIR_2005}.
With the rapid growth of spintronics, the spin properties of Hg$_{1-x}$Cd$_{x}$Te systems have attracted more and more attention in recent decades. Various spin-related phenomena have been discovered in HgTe-based quantum wells (HgTe QWs), such as giant \cite{ZhangXC_HgTeRSS_2001,GuiYS_GiantSS_2004} and
nonlinear spin splitting \cite{Wyang_NonlinearRSS_2006,LiuXZ_NoninearRSS_2013}, large effective $g$ factor \cite{ZhangXC_HgTegFactor_2004}, and intrinsic spin Hall effect \cite{Wyang_ISHE_2008,Brune_ISHE_2008}.
At the heart of these spin-related phenomena, the very strong spin-orbit interaction (SOI) of Hg$_{1-x}$Cd$_{x}$Te plays an essential role.
Moreover, the two-dimensional(2D) topological insulator (TI) phase emerges because the strong SOI could drive HgTe QWs into the inverted-band regime \cite{Bernevig_QSHE_2006,Konig_QSHE_2007}. The strong SOI comes from the large relativistic corrections of heavy atoms Hg, Cd, and Te, which makes Hg$_{1-x}$Cd$_{x}$Te-based systems become ideal platforms for the study of spintronics, topological electronics, as well as the spin-resolved infrared and terahertz optoelectronics\cite{MacDonald_Spintronics_2012,QiXiaoLiang_TIReview_2011,LBZhang_HgTePNJ_2010,Kchang_HgTeQDot_2011,Ganichev_THzPLDetect_2008,Ganichev_THzPLDetect_2008,Danilov_THZdetecor_2009}.

Circular photogalvanic effect (CPGE), which is identified by the direction reverse of photocurrents when changing the helicity of circularly polarized light, has been intensively studied in semiconductors \cite{Ivchenko1997, Ganichev_PRLCPGE_2001,Ganichev_CPGERev_2003,Diehl_NJP_2007}. Microscopically, CPGE is caused by the conversion of photon angular momentum into translational motion of carriers and is sensitively dependent on the zero-field spin splitting. In low-dimensional semiconductors, the zero-field spin splitting can be ascribed to two different kinds of SOI terms, i.e., the Rashba SOI (RSOI) term comes from the structure inversion asymmetry (SIA) \cite{Rashba_RSOI_1960}, and the Dresselhaus SOI (DSOI) term originates from the crystal bulk inversion asymmetry (BIA) \cite{Dresselhaus_DSOI_1955}. Therefore, CPGE actually forms a bridge between the photocurrent signals and the symmetry and SOI information of host materials. In various semiconductor systems, such as low-dimensional structures of GaAs, InAs, SiGe, GaN and ZnO, CPGE has been successfully used as a tool to determine the relative ratio of Rashba and Dresselhaus terms (RD ratio) \cite{Ganichev_RDratio_2004,Giglberger_RDration_2007, Ganiche_SiGeCPGE_2002, QZhang_ZnOCPGE_2010, CMYin_GaNCPGE_2010, CYH_RDQWrs_2011, Ganichev_RDinterplay_2014}. Because of the unique SOI property and novel TI phase of HgTe QWs, CPGE in HgTe QWs has also attracted considerable interest \cite{Wittman_HgTeCPGE_2010,Olbrich_PhotoCurrentCR_2013,Kaladzhyan_BulkedgeCPGE_2015}. Experimentally, large CPGE signals in (001)- and (113)-oriented HgTe QWs have been observed in terahertz and mid-infrared regions \cite{Wittman_HgTeCPGE_2010,Olbrich_PhotoCurrentCR_2013}, and have found their application in the fast detection of the infrared-radiation ellipticity \cite{Ganichev_THzPLDetect_2008,Danilov_THZdetecor_2009}.

In conventional semiconductors, the microscopical picture of CPGE can be well described by $\mathbf{k}$-linear Rashba and Dresselhaus models \cite{Ganichev_CPGERev_2003,Golub_CPGE_2003,ZhouBin_CPGE_2007}.
However, due to the narrow gap and strong SOI, the band structures and spin splittings of HgTe QWs are very distinct from those of conventional semiconductor systems \cite{Ortner_VBStructureHgTe_2002,Novik_HgMnTeBand_2005,Wyang_NonlinearRSS_2006}.
More importantly, there is a substantial change of band structure when HgTe QWs undergo the TI phase transition \cite{Bernevig_QSHE_2006,JunLi_EDrivePhase_2009}.
As a consequence, the experimental CPGE signals were found to be about an order of magnitude larger than those observed in conventional semiconductor QWs \cite{Wittman_HgTeCPGE_2010}.
This suggests that theory based on detailed band model beyond previous reduced models is required to study the unique CPGE in HgTe QWs.
In this paper, we present a theoretical method to calculate CPGE photocurrents based on the eight-band $\mathbf{k\cdot p}$ model combing with density-matrix formalism.
Using this method, we investigate the microscopic origin and the pseudotensor of CPGE of coefficients in HgTe QWs ranging from the normal to heavily inverted regime.
We find the CPGE could be significantly enhanced in heavily inverted HgTe QWs, which is consistent with the large photocurrent signals observed in Ref. \onlinecite{Wittman_HgTeCPGE_2010}. This enhancement effect could be utilized as an experimental evidence of HgTe QWs in heavily inverted phase and could provide advantages in improving the efficiency of the ellipticity detector of the infrared and terahertz radiation \cite{Danilov_THZdetecor_2009}. In addition, by adding the eight-band BIA terms, in Sec. \ref{sec:BIAeffect} we also discuss the interplay effect of BIA and SIA on CPGE currents. And within the same theoretical framework, in Sec. \ref{sec:PSC} we investigate the pure spin currents (PSCs) generated by linearly polarized light under normal incidence in HgTe QWs. An interesting finding is that the pure spin current $j_{y'}^{x'}$ ($x'\parallel[110]$ and $y'\parallel[\bar{1}10]$) driven by $[110]$ linearly polarized light changes sign when HgTe QWs are transformed from normal phase to inverted phase.

\section{Band-structure Model and Theoretical formalism \label{sec:theory}}

\begin{figure}[tpb]
\centering
\includegraphics[width=0.85\columnwidth]{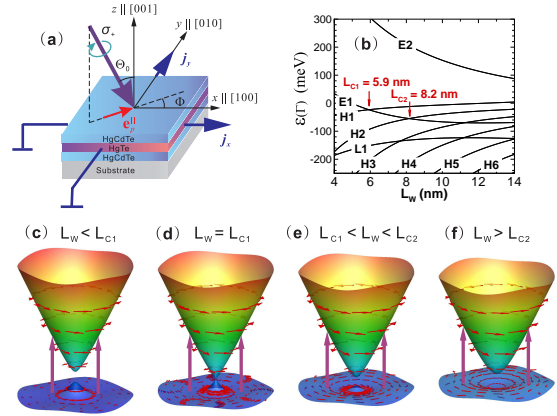}% Here is how to import EPS art
\centering{\caption[c]{\label{fig1} (Color online) (a) Schematic of the right-handed circularly polarized ($\sigma^{+}$) light irradiating on the HgTe QW. The CPGE current is detected along the $[100]$ and $[010]$ crystallographic directions. The red arrow denotes the projection of the light propagation direction unit vector on the QW plane, i.e., $\mathbf{e}_{p}^{\parallel}$. (b) Subband energies at the $\Gamma$ point of the Brillouin zone as a function of well width $\mathrm{L}_\mathrm{w}$. (c)-(f) The 3D band structures and spin textures (red arrows) for HgTe QWs with different $\mathrm{L}_\mathrm{w}$: 5, 5.9, 7 and 9 nm, corresponding to HgTe QW in normal, Dirac-like, inverted, and heavily inverted regimes. The up purple arrows indicate the direct optical transitions.}}
\end{figure}

As sketched in Fig. \ref{fig1}(a), in this work we consider a (001)-grown strain-free HgTe QW with Hg$_{0.3}$Cd$_{0.7}$Te as barriers and the $x$, $y$, $z$ axis aligned with the $[100]$, $[010]$ and $[001]$ crystallographic orientations, respectively. The spin splitting is an indispensable ingredient for the generation of CPGE \cite{Golub_CPGE_2003}. So the Hamiltonian of HgTe QWs, i.e, $\hat{H}_{0}$, should contain a spatial inversion symmetry and asymmetry part
\begin{equation}
\hat{H}_{0}=\hat{H}_{K}+\hat{H}_{A},
\end{equation}%
where $\hat{H}_{K}$ and $\hat{H}_{A}$ represent the inversion symmetry and asymmetry part, respectively. In this paper, $\hat{H}_{K}$ is taken as the modified eight-band Kane Hamiltonian of a symmetric HgTe QW (see Appendix \ref{Apdx:Hamiltonian}), which does not produce the spin splitting. The source of spin splitting comes from $\hat{H}_{A}$, which could either be SIA or BIA. In HgTe-based QWs, SIA is found to be the dominant mechanism of spin splitting \cite{ZhangXC_HgTeRSS_2001}. So in Secs. \ref{sec:CPGE_Origin} and \ref{sec:CPGE_SIA}, we will focus on the SIA-induced CPGE  and the influence of BIA will be considered in Sec. \ref{sec:BIAeffect}.

The bulk HgTe is a semimetal with negative gap. If HgTe is sandwiched in between two barriers to form a QW, the band gap could be tuned from the negative regime to positive regime by quantum confinement effect. With increasing the thickness of the HgTe layer, i.e., the well width $\mathrm{L}_\mathrm{w}$, a TI phase transition takes place at the critical thickness $\mathrm{L}_\mathrm{c1}=$ 5.9 nm as shown in Fig. \ref{fig1}(b). If $\mathrm{L}_\mathrm{w}<\mathrm{L}_\mathrm{c1}$, the E1 subband lies above the H1 subband at the $\Gamma$ point like a normal semiconductor, corresponding to the band insulator (BI) phase of HgTe QWs, though in the BI phase, the conduction bands of HgTe QWs show non parabolic behavior which is different from the wide-gap semiconductors [see Fig. \ref{fig1}(c)]. If $\mathrm{L}_\mathrm{w}>\mathrm{L}_\mathrm{c1}$, the order of E1 and H1 is inverted, i.e., H1 lies above E1, so that the HgTe QW is in inverted phase [see Fig. \ref{fig1}(e)]. In this phase, there will be a pair of robust spin-momentum-locked states counterpropagating at the edges of the finite QW plane, which could lead to the quantum spin Hall effect \cite{Bernevig_QSHE_2006,Konig_QSHE_2007}. This phase is also referred to as the 2D TI phase and has attracted extensive enthusiasms. At the critical thickness, i.e., $\mathrm{L}_\mathrm{w}=\mathrm{L}_\mathrm{c1}$, the low-energy band dispersion of HgTe QWs is like a Dirac-cone with zero gap [see Fig. \ref{fig1}(d)], which is a promising system for the study of Dirac fermion physics \cite{Buttner_DiracHgTe_2011}. If one further increase $\mathrm{L}_\mathrm{w}$ to reach  $\mathrm{L}_\mathrm{w}>\mathrm{L}_\mathrm{c2}=$ 8.2 nm, HgTe QWs could enter the heavily inverted regime so that E1 even falls below H2, and H1 (H2) becomes the first conduction (valence) subband. It is thus of great interest to investigate how the CPGE evolves when HgTe QW undergoes the quantum phase transition.

Consider that the HgTe QW is irradiated by a beam of
single-color polarized light with frequency $\omega$ (in the terahertz to infrared region). The incident angle and azimuthal angle of light is denoted by $\Theta _{0}$ and $\Phi$, respectively [as sketched in Fig. \ref{fig1}(a)].
The total Hamiltonian can be written as
\begin{equation}
\centering
\hat{H}=\hat{H}_{0}+\hat{V}(t),
\end{equation}%
where $\hat{V}(t)$ is the electron-radiation interaction.
\begin{equation}
\hat{V}(t)=\hat{S}e^{-i\omega t}+\hat{S}^{\dag }e^{i\omega t}.
\end{equation}
and
\begin{equation}
\hat{S}\equiv \frac{e}{i m_{0} \omega}\mathbf{E}\cdot \hat{\mathbf{p}},
\end{equation}
Here $\hat{\mathbf{p}}=m_{0}\hat{\mathbf{v}}$ is the momentum operator. $m_{0}$ is the free electron mass, and $\hat{\mathbf{v}}$ is the velocity vector operator. In the representation of eight-band basis [Eq. (\ref{eqn:basis})], the three components of the velocity operator, i.e., $\hat{v}_{x}$, $\hat{v}_{y}$ and $\hat{v}_{z}$ are an eight-by-eight matrix, and can be derived by \cite{Wyang_ISHE_2008}
\begin{equation}\label{velocityop}
\hat{v}_{\alpha}=\frac{1}{i\hbar }[\hat{r}_{\alpha }, \hat{H}_{0}], \quad (\alpha\in \{x,y,z\}).
\end{equation}
$\mathbf{E}$ is the complex amplitude of the light electric field.
The three components of $\mathbf{E}$ can be written as
\begin{eqnarray}\label{LEvertor}
E_{x}&=&\frac{E_{0}}{\sqrt{2}}\left( t_{p}\cos \Theta \cos \Phi e^{i\varphi
}-t_{s}\sin \Phi e^{-i\varphi }\right) ,\nonumber \\
E_{y}&=&\frac{E_{0}}{\sqrt{2}}\left( t_{p}\cos \Theta \sin \Phi e^{i\varphi
}+t_{s}\cos \Phi e^{-i\varphi }\right) ,\nonumber \\
E_{z}&=&\frac{E_{0}}{\sqrt{2}}t_{p}\sin \Theta e^{i\varphi }.
\end{eqnarray}
In Eq. (\ref{LEvertor}), $E_{0}$ is the electric-field amplitude in vacuum.
$\Theta $ is the refraction angle determined by $\sin \Theta =\sin \Theta _{0}/n_r$.
$\varphi $ is half of the phase angle between the two perpendicular components of the light electric field.
$t_{p}$ and $t_{s}$ are the transmission coefficients for the $p$ and $s$
polarization components of the light electric field.
$E_{0}$ is dependent on the intensity of light via $E_{0}=\sqrt{2I_{0}/(c_{0}n_r\varepsilon _{0}})$, where
$I_{0}$, $c_{0}$, $\varepsilon _{0}$ and $n_r$ are the intensity of light, light speed in vacuum,
dielectric constant in vacuum, and the refraction index of QWs, respectively.
$t_{p}$ and $t_{s}$ can be found by Fresnel's formula:
$t_{p}=2\cos \Theta /(n\cos \Theta +\sqrt{1-\sin
^{2}\Theta /n^{2}})$, $t_{s}=2\cos \Theta /(\cos \Theta +\sqrt{n^{2}-\sin
^{2}\Theta })$. Using Eq. (\ref{LEvertor}), one can verify $i\mathbf{E}\times
\mathbf{E}^{\ast }\varpropto t_{p}t_{s}P_{circ}\left\vert E_{0}\right\vert
^{2}\mathbf{e}_{p}$, where $P_{circ}\equiv \left( I_{\sigma +}-I_{\sigma
-}\right) /\left( I_{\sigma +}-I_{\sigma -}\right) =\sin 2\varphi $ is the helicity of the incident light,
and $\mathbf{e}_{p}$ is the unit vector of the light propagation direction.
Changing $\varphi $ from $45$ to $135^\circ$, the incident light could be continuously varied from right-handed circularly polarized ($\sigma ^{+}$) to left-handed circularly polarized ($\sigma^{-}$).
For the $\sigma ^{+}$ ($\sigma^{-}$) light, its angular momentum has the nonzero in-plane component parallel (antiparallel) to the projection of $\mathbf{e}_{p}$ on the QW plane, as denoted by $\mathbf{e}_{p}^{\parallel}$ in Fig. \ref{fig1}(a).

Density-matrix formalism provides a quantum-mechanics approach for the microscopic description of the linear and non-linear optical susceptibilities \cite{NonlinearOptics3Ed}.
Following this formalism, other optical quantities, such as circular photogalvanic currents and linear photogalvanic pure spin currents, can also be calculated. We shall start from the Liouville equation which describes the time evolution of the density matrix.
Using the eigenstates of $\hat{H}_{0}$ as the basis set, i.e., $\{|m,\mathbf{k}\rangle\}$, the Liouville equation can be written as
\begin{equation}\label{LVeq}
\frac{\partial \rho _{mn}}{\partial t}=-\frac{i}{%
\hbar }[\hat{H},\hat{\rho}] _{mn}-\Gamma _{mn}[\rho
_{mn} -\rho _{mn}^{eq}].
\end{equation}%
In Eq. (\ref{LVeq}), $\hat{\rho}$ is the density operator, and we have used the notation $A_{mn}=A_{mn}(\mathbf{k})\equiv\langle m,\mathbf{k}|\hat{H}|n,\mathbf{k}\rangle$ for the matrix elements of operator $\hat{A}$. $\rho _{mn}^{eq}$ is the initial density matrix. At thermal equilibrium, $\rho _{mn}^{eq} = f_{m}\delta _{mn}$, where $f_{m}$ is the Fermi distribution function.
The second term on the right-hand side of Eq. (\ref{LVeq}) is a phenomenological damping term.
$\Gamma _{nn}$ represents the decay rate for the nonequilibrium carriers in the $n$th subband, and $\Gamma _{mn}^{(m\neq n)}$ describes the dephasing rate of $\rho _{mn}$ coherence. In this paper, we take $\Gamma _{nn}=1/T_{1}$ with $T_{1}=$ 200 ps as a typical recombination lifetime of the direct gap semiconductor. And $\Gamma _{mn}^{(m\neq n)}=1/T_{2}$ with $T_{2}=$ 1.3 ps, which is a reasonable dephasing time of $\rho _{mn}$ in semiconductors at room temperature, and could cause a 1-meV level broadening in spectra \cite{Klingshirn2012}.

By treating $\hat{V}(t)$ as the perturbation, and expanding $\hat{\rho}$ as the sum of the zeroth-, first-,
second-order components: $\hat{\rho}\approx \hat{\rho}^{(0)} +\hat{\rho}^{(1)}+\hat{\rho}^{(2)}$, we have
\begin{eqnarray} \label{rhoex}
\frac{\partial \rho _{mn}^{(0) }}{\partial t}&=&-\frac{i}{\hbar
}[\hat{H}_{0},\hat{\rho}^{( 0) }] _{mn}-\Gamma
_{mn}(\rho _{mn}^{( 0) }-\rho _{mn}^{eq}) ,  \nonumber \\
\frac{\partial \rho _{mn}^{( 1) }}{\partial t} &=&-\frac{i}{\hbar
}[ \hat{H}_{0},\hat{\rho}^{( 1) }] _{mn}-\frac{i}{%
\hbar }[ \hat{V}( t) ,\hat{\rho}^{( 0) }]
_{mn}-\Gamma _{mn}\rho _{mn}^{( 1) },  \nonumber \\
\frac{\partial \rho _{mn}^{( 2) }}{\partial t} &=&-\frac{i}{\hbar
}[ \hat{H}_{0},\hat{\rho}^{( 2) }] _{mn}-\frac{i}{%
\hbar }[ \hat{V}( t) ,\hat{\rho}^{( 1) }]
_{mn}-\Gamma _{mn}\rho _{mn}^{(2)}.
\end{eqnarray}
We are interested in the second-order steady-state solution of Eq. (\ref{rhoex}), i.e., $\rho_{mn}^{(2)}( t)$ with $ t\rightarrow\infty$, which is found to be
\begin{eqnarray} \label{rho2}
\rho_{mn}^{(2)\infty}(\mathbf{k}) & \equiv & \rho_{mn}^{(2)}( t\rightarrow\infty)  \nonumber \\
 & = & -\frac{1}{\hbar^{2}(\omega_{mn}-i\Gamma _{mn})} %\nonumber \\
 \times \sum_{q} \{ S_{mq}(S^{\dag })_{qn}( \frac{f_{q}-f_{n}}{\omega _{qn}+\omega
-i\Gamma _{qn}}+\frac{f_{q}-f_{m}}{\omega _{mq}-\omega -i\Gamma _{mq}}) \nonumber \\
& & +(S^{\dag})_{mq}S_{qn}( \frac{f_{q}-f_{m}}{\omega _{mq}+\omega
-i\Gamma _{mq}}+\frac{f_{q}-f_{n}}{\omega _{qn}-\omega -i\Gamma _{qn}})\}.
\end{eqnarray}%
Here $\omega_{mn}\equiv [\varepsilon_m(\mathbf{k})-\varepsilon_n(\mathbf{k})]/\hbar$, and $\varepsilon_m(\mathbf{k})$ is the eigenenergy of $\hat{H}_{0}$. $S_{mn}$ and $(S^{\dag})_{mn}$ are the matrix elements of $\hat{S}$ and $\hat{S}^{\dag}$. In principle, $m,n,q$ should run over all subbands of HgTe QWs.
However, for the absorption of single-color light, only a few subbands (less than ten subbands) with energy differences in the range of photon energy needs to be taken into account, because optical process obeys energy conservation law. After obtaining the set of eigenenergies and eigenstates of HgTe QWs, i.e., $\{\varepsilon_m(\mathbf{k})\}$, $\{|m,\mathbf{k}\rangle\}$ by solving Eq. (\ref{eq:sch}), $\rho_{mn}^{(2)\infty}( \mathbf{k})$ can be calculated according to Eq. (\ref{rho2}).

The density matrix can be used to evaluate the expectation value of an arbitrary operator $\hat{%
O}$
\begin{equation} \label{expvalue}
\langle\hat{O}\rangle =\sum_{\mathbf{k}}Tr\{\hat{O}\hat{\rho}\}=\sum_{\mathbf{k},m,n}\rho
_{mn}( \mathbf{k}) \langle n,\mathbf{k}\vert \hat{O}%
\vert m,\mathbf{k}\rangle .
\end{equation}
Equation (\ref{expvalue}) contains a Brillouin-zone integration over the in-plane $\mathbf{k}$ space.
In order to achieve numerical results with acquired accuracy, over 150 000 states on in-plane $\mathbf{k}$ grids needs to be calculated.
For each $\mathbf{k}$, a $(16N+8)$-by-$(16N+8)$ Hamiltonian matrix generated by plane-wave expansion method [see Eq.(\ref{eq:pwex})] is required to be diagonalized.
A parallel numerical program based on Message Passing Interface (MPI) is designed to accelerate this calculation. Note that in contrast to the previous works \cite{Ganichev_CPGERev_2003,Golub_CPGE_2003,ZhouBin_CPGE_2007},
the calculation by Eq. (\ref{expvalue}) takes account of a finite energy level broadening through $\hbar \Gamma _{mn}^{(m\neq n)}$ in $\rho_{mn}^{(2)\infty}$.
From Eq. (\ref{LVeq}), one can find the broadening has a clear physical origin related to the dephasing of $\rho_{mn}$, which can be caused by the collisions of atoms at finite temperature \cite{NonlinearOptics3Ed}.
The broadening allows the off-diagonal part of $\rho_{mn}^{(2)\infty}$, i.e., $\rho_{mn}^{(2)\infty}$ with $m\neq n$, to be nonzero by the excitation of a single-color light with fixed frequency.
Therefore, the coherent terms such as $\rho_{mn}O_{nm}$ with $m\neq n$ can be rigorously taken into account in our calculation, while they were usually neglected or treated approximately in the previous works.
In addition, the broadening of energy levels leads to a finite peak-like integrand in Eq. (\ref{expvalue}), which can be directly calculated by using the standard numerical quadrature procedure, such as the Gaussian quadrature.

\section{Numerical Results and discussions of CPGE}

\subsection{Microscopic origin of CPGE \label{sec:CPGE_Origin}}

In this section, we will discuss the microscopic origin of CPGE and show its relation to SOI.
Applying Eq. (\ref{expvalue}), the photogalvanic charge currents can be calculated by
\begin{equation} \label{eq:J1}
\mathbf{j}=-e\sum_{\mathbf{k},m,n}\rho _{mn}( \mathbf{k}) \mathbf{%
v}_{nm}( \mathbf{k}).
\end{equation}
Because the Hamiltonian of HgTe QWs is a time-reversal invariant,
it guarantees the Kramers' degeneracy, i.e., $\varepsilon _{n}( \mathbf{k})
=\varepsilon _{\bar{n}}( -\mathbf{k})$. Here and in the following, a bar
above the subband index denotes the subband with the opposite spin. By applying the
time reversal operation, we can demonstrate $\mathbf{v}_{mn}( \mathbf{k}%
) =-\mathbf{v}_{\bar{n}\bar{m}}^{\ast }( -\mathbf{k}) =-%
\mathbf{v}_{\bar{m}\bar{n}}( -\mathbf{k})$. Using these
properties, Eq. (\ref{eq:J1}) can be split as
\begin{eqnarray}\label{eq:J2}
\mathbf{j} &=&-\frac{e}{2}\sum_{\mathbf{k},m,n}[ \rho _{mn}( \mathbf{k}%
) \mathbf{v}_{nm}( \mathbf{k}) +\rho _{mn}( -\mathbf{k}%
) \mathbf{v}_{nm}( -\mathbf{k}) ]  \nonumber \\
&=&-\frac{e}{2}\sum_{\mathbf{k},m,n}[ \rho _{mn}( \mathbf{k}%
) \mathbf{v}_{nm}( \mathbf{k}) -\rho _{mn}( -\mathbf{k}%
) \mathbf{v}_{\bar{n}\bar{m}}( \mathbf{k}) ].
\end{eqnarray}
Next, we can change the dummy subscripts $m,n$ of the second term by $\bar{m},\bar{n}$,
and get
\begin{equation} \label{eq:J3}
\mathbf{j}=-\frac{e}{2}\sum_{\mathbf{k,}m,n}[ \rho _{mn}( \mathbf{k%
}) -\rho _{\bar{m}\bar{n}}( -\mathbf{k}) ] \mathbf{v}%
_{nm}( \mathbf{k}).
\end{equation}
Equation (\ref{eq:J3}) demonstrates that the photocurrents come from the
nonsymmetrical distribution of photoexcited density matrix at $\mathbf{k}$
and $-\mathbf{k}$ points. We can describe the asymmetrical part of the density matrix
by $\Delta \rho _{mn}(\mathbf{k})\equiv \rho_{mn}(\mathbf{k})-\rho _{\bar{m}\bar{n}}(-\mathbf{k})$.
In steady condition, only the second-order density matrix contributes to the asymmetry, which gives $\Delta \rho _{mn}(\mathbf{k})=\rho _{mn}^{(2)\infty}(\mathbf{k})-\rho _{\bar{m}\bar{n}}^{(2)\infty}(-\mathbf{k})$.
Due to the spin-dependent selection rule, the circularly polarized light would give rise to different transition rates for
$|m(n),\mathbf{k}\rangle\leftrightarrow |q,\mathbf{k}\rangle $ from $|\bar{m}(\bar{n}),-\mathbf{k}\rangle \leftrightarrow |\bar{q},-\mathbf{k}\rangle $, thus breaking the symmetry of $\rho _{mn}(\mathbf{k})$ [or causing $\Delta \rho _{mn}(\mathbf{k})\neq$ 0] according to Eq. (\ref{rho2}). As a consequence, a net charge current would emerge along the asymmetrical direction of $\rho _{mn}(\mathbf{k})$. This process is equivalent to the transformation of photon angular momenta into translational motion of free carriers \cite{Ganichev_PRLCPGE_2001}.

\begin{figure} [tpb]
\includegraphics[width=0.85\columnwidth]{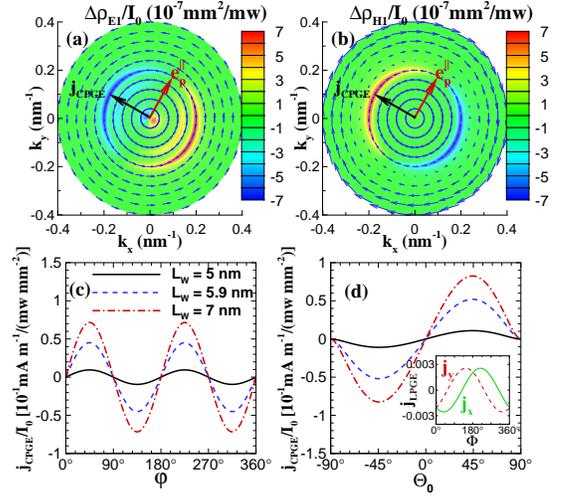}
\caption{\label{fig2}(Color online) (a) and (b) Calculated $\bm{\mathfrak{B}}_n(\mathbf{k})$ of RSOI (minor blue arrows) and $\Delta \rho_{nn}(\mathbf{k})$ (contour color) of a 5.9 nm HgTe QW for $n=$ E1 and H1 subband respectively. They are induced by a $\sigma^{+}$ light with $\hbar\omega=$ 129 meV, $\Theta _{0}=$ 30$^\circ$ and $\Phi=$ 60$^\circ$. The red arrows indicate the projection of light propagation direction in the QW plane and the black arrows indicate the direction of CPGE current. (c) and (d) The magnitudes of photocurrent as a function of half phase angle $\varphi$ ($\Theta _{0}=$ 30$^\circ$)
and the incident angle of light $\Theta _{0}$ ($\varphi=$ 45$^\circ$), for HgTe QWs with different QW widths. The inset of (d) shows the components $j_{x}$ and $j_{y}$ of LPGE current as a function of $\Phi$ for a 5.9 nm HgTe QW ($\varphi=$ 0$^\circ$ and $\Theta _{0}=$ 30$^\circ$). The unit of LPGE current is the same as that of photocurrents in (c) and (d).}
\end{figure}

In order to manifest the role of SOI in the generation of CPGE, we introduce the notion of effective magnetic field of SOI \cite{Winkler2003} [denoted by $\bm{\mathfrak{B}}_n(\mathbf{k})$ as defined in Appendix \ref{Apdx:EMF}]. $\bm{\mathfrak{B}}_n(\mathbf{k})$ can be regarded as the effective magnetic field felt by an electron with state $|n,\mathbf{k}\rangle$ due to spin-orbit coupling. The magnitude and direction of $\bm{\mathfrak{B}}_n(\mathbf{k})$ can describe the SOI spin splitting of $n$th subband and the spin orientation of the upper spin branch, respectively.
In Figs. \ref{fig2}(a) and \ref{fig2}(b), we plot the calculated effective magnetic fields $\bm{\mathfrak{B}}_n(\mathbf{k})$ of RSOI and the asymmetrical parts of diagonal density-matrix elements, i.e., $\Delta \rho_{nn}(\mathbf{k})$, for $n=$ E1 and H1, respectively.
The picture of $\Delta \rho_{nn}(\mathbf{k})$ can be viewed as the $\mathbf{k}$-space distribution of nonequilibrium carrier density of the $n$th subband. The $\sigma^{+}$ light at oblique incidence gives a non-zero in-plane angular momentum component along $\mathbf{e}_{p}^{\parallel}$, so it will excite more E1 (H1) states with $\bm{\mathfrak{B}}_n(\mathbf{k})$ parallel (antiparallel) to $\mathbf{e}_p^{\parallel}$. For HgTe QWs with SIA only, the systems hold $C_{4v}$ point group symmetry, restricting $\bm{\mathfrak{B}}_n(\mathbf{k})$ perpendicular to $\mathbf{k}$ for most E1 and H1 states.
Therefore we can see the maximum of $\Delta\rho_{nn}(\mathbf{k})$ will appear at the direction perpendicular to $\mathbf{e}_p^{\parallel}$ in $\mathbf{k}$ space. Note that though E1 and H1 states have opposite signs of $\Delta \rho_{nn}(\mathbf{k})$, their effective masses and velocities are also opposite in signs, so they have the same direction contribution to the net charge currents perpendicular to $\mathbf{e}_p^{\parallel}$, instead of canceling each other. In Fig. \ref{fig2}(c), we can see the calculated photocurrents clearly exhibit the signature of CPGE, i.e, the sign dependency of light's helicity. Fig. \ref{fig2}(d) shows the photocurrents as a function of incident angle $\Theta _{0}$. The most effective incident angle to generate CPGE is $\Theta _{0}= \pm 45^\circ$, because at this incident angle, the refraction light has the largest in-plane angular momentum component, as determined by the Fresnel's formula. In general, we find the photocurrents increase with the width of the well. This is because the wider QWs have larger SIA spin splitting under the same magnitude of electric field.

Phenomenologically, the photogalvanic currents can be described by \cite{Ivchenko1997}
\begin{equation}  \label{eq:J4}
j_{\lambda }=\sum_{\alpha, \beta }\chi _{\lambda \alpha \beta }E_{\alpha
}E_{\beta }^{\ast },
\end{equation}%
where $\chi _{\lambda \alpha \beta }$ is a third-rank phenomenological
tensor. For the in-plane photocurrents, $\lambda\in\{x,y\}$ and $\alpha, \beta \in\{{x,y,z}\}$. Inserting  Eq. (\ref{rho2}) into Eq. (\ref{eq:J1}) and comparing with Eq. (\ref{eq:J4}), we can find the microscopic expression for $\chi _{\lambda \alpha \beta }$ is
\begin{eqnarray}\label{chi}
\chi _{\lambda \alpha \beta }  =  \frac{e^{3}}{\omega ^{2}\hbar ^{2}}\sum_{\mathbf{k}}\sum_{mnq}( f_{q}-f_{m})\times[v_{mq}^{\alpha}v_{qn}^{\beta}v_{nm}^{\lambda }\mathcal{L}_{mnq}(\omega) + v_{nq}^{\alpha }v_{qm}^{\beta }v_{mn}^{\lambda }\mathcal{L}^{\ast }_{mnq}(\omega ) \nonumber \\
+v_{qn}^{\alpha}v_{mq}^{\beta}v_{nm}^{\lambda}\mathcal{L}_{mnq}(-\omega )+v_{qm}^{\alpha }v_{nq}^{\beta }v_{mn}^{\lambda }\mathcal{L}^{\ast }_{mnq}(-\omega )],
\end{eqnarray}
where $\mathcal{L}_{mnq}(\omega) \equiv 1/[( \omega _{mn}-i\Gamma
_{mn}) ( \omega _{mq}-\omega -i\Gamma _{mq})].$ From Eq. (\ref{chi}),
we can verify $\chi _{\lambda \alpha \beta }=\chi _{\lambda
\beta \alpha }^{\ast }$. This property allows one to decompose Eq. (\ref{eq:J4}) into two terms by the symmetric and anti-symmetric sum of $E_{\alpha }E_{\beta }^{\ast }$, respectively:
\begin{equation} \label{eq:J5}
j_{\lambda }=\sum_{\alpha \beta }Re(\chi _{\lambda \alpha \beta })
\frac{E_{\alpha }E_{\beta }^{\ast }+E_{\alpha }E_{\beta }^{\ast }}{2} \\
+i\sum_{\alpha \beta }Im(\chi _{\lambda \alpha \beta })\frac{E_{\alpha
}E_{\beta }^{\ast }-E_{\alpha }^{\ast }E_{\beta }}{2}.
\end{equation}
The first and second term on the right-hand side of Eq. (\ref{eq:J5}) describe the linear photogalvanic effect and circular photogalvanic effect, or LPGE and CPGE,
respectively. The LPGE is only allowed in noncentrosymmetric crystals of the piezoelectric classes \cite{Ivchenko1997,Ganichev_CPGERev_2003}. In (001)-oriented HgTe QWs, experiments show the LPGE currents are small compared to the CPGE currents \cite{Wittman_HgTeCPGE_2010}.
In our calculation, we also find the LPGE currents are two orders of magnitude smaller than CPGE currents in (001)-oriented HgTe QWs. This can be seen in the inset of Fig. \ref{fig2}(d).
Therefore in this paper, we can simply neglect the LPGE term, and concentrate on the CPGE term, which can be rewritten as a commonly referred form \cite{Ganichev_CPGERev_2003,Wittman_HgTeCPGE_2010}
\begin{equation}\label{eq:J6}
j_{\lambda}=\sum_{\mu}\gamma _{\lambda \mu }i( \mathbf{E\times E}%
^{\ast }) _{\mu},
\end{equation}
where $\gamma _{\lambda \mu }$ ($\lambda,\mu\in\{x,y\}$) is a second-rank pseudotensor. Using the
Levi-Civita antisymmetric tensor $\varepsilon _{\alpha \beta \lambda }$,
we can write $\gamma _{\lambda \mu }=\mathrm{Im}(\chi _{\lambda \alpha \beta
})\varepsilon _{\alpha \beta \mu }$. However, we should mention that both LPGE and CPGE could have significant contributions to the photocurrents in HgTe QWs with low symmetries, such as the QWs grown on high-index-planes \cite{Wittman_HgTeCPGE_2010}.

\subsection{CPGE induced by structure inversion asymmetry \label{sec:CPGE_SIA}}
In low-dimensional semiconductors, SIA may arise from the asymmetrical heterostructure materials, confining potentials or dopings, as well as the external or built-in electric fields. For the single conduction band model, SIA is usually described by a $\mathbf{k}$-linear term, or so-called Rashba term. To study the SIA effect, in the framework of the eight-band $\mathbf{k\cdot p}$ model, one can simply introduce a static electric field $\mathcal{F}$ along the $z$ axis, i.e., let
\begin{equation} \label{HSIA}
\hat{H}_{A}=\hat{H}_{SIA}=-e\mathcal{F}z.
\end{equation}
As demonstrated by Pfeffer and Zawadzki, the SIA spin splitting is dominated by the asymmetry of overlap between the valence-band offset and the electron's envelope function at the interfaces \cite{Pfeffer_Spinsplitting_1999}.
The effect of Eq. (\ref{HSIA}) is to make the electron envelope function asymmetric, so that its overlap with the valence band offset at the interfaces also becomes asymmetric and hence the SIA spin splitting is produced.
Because the asymmetry of electron's envelope function can be effectively tuned by the magnitude of $\mathcal{F}$ \cite{Wyang_RSSanalysis_2006}, it forms the fundament of manipulating the spin splitting and spin states with external electric field. Different from the conventional $\mathbf{k}$-linear Rashba term, the spin splitting in HgTe QWs produced by $\hat{H}_{SIA}$ is nonlinear in $\mathbf{k}$, because the kinetic energy of electrons is comparable to the narrow band gap \cite{Wyang_NonlinearRSS_2006,LiuXZ_NoninearRSS_2013}. This is a general feature of narrow-gap systems, and here it can be exactly taken into account by the eight-band model. $\hat{H}_{SIA}$ also leads to a $C_{4v}$-symmetric $\bm{\mathfrak{B}}_n(\mathbf{k})$ or spin textures, which requires $\gamma _{yx}=-\gamma_{xy}$ be the only nonzero components of $\gamma _{\lambda \mu }$.

\begin{figure}[tpb]
\includegraphics[width=0.85\columnwidth]{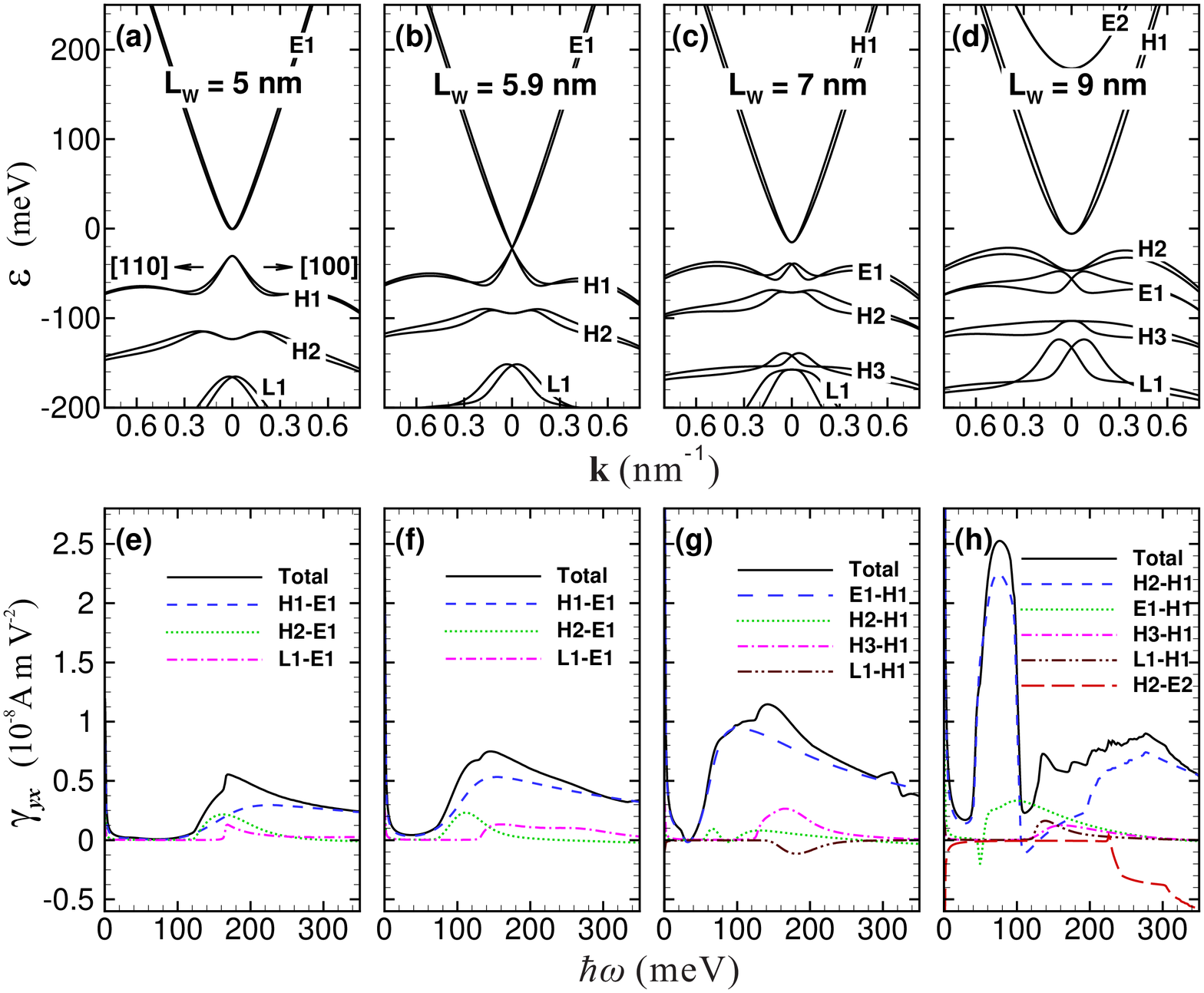}% Here is how to import EPS art
\caption{\label{fig3}(Color online) (a)-(c) The band structures of HgTe QWs with SIA ($\mathcal{F}=$ 80 kV/cm),
for $\mathrm{L}_\mathrm{w} =$ 5, 5.9, 7, and 9 nm respectively.
(d)-(f) The calculated spectra of $\gamma _{xy}$ with respect to (a)-(c).
The black solid lines are the total spectrums of $\gamma_{xy}$ including all contributions, and the other lines are contributions from each subband-subband transitions.}
\end{figure}

In Fig. \ref{fig3}, we present the calculated band structures for HgTe QWs
in different regimes, the corresponding spectra of $\gamma _{yx}$, and contributions from each subband-subband transitions. Generally, we can see the spectra of $\gamma_{yx}$ are sensitively dependent on the band structures and spin splitting. And the largest contribution to $\gamma_{yx}$ come from the transition of the top valence subband to the bottom conduction subband. In wider HgTe QWs, there are more subbands involved in the optical transition and larger Rashba spin splitting, so the maximum of $\gamma_{yx}$ increases with $\mathrm{L}_\mathrm{w}$. Note that because we consider an infinite HgTe quantum well with no edges or boundaries, the edge states do not show up in the bulk gap of inverted HgTe QWs [as can be seen in Fig. \ref{fig3}(g) and (h)]. However, if the frequency is larger than the bulk band gap, we would expect the optical transitions between bulk subbands to dominate, since the bulk states have a much larger density of states compared with the edge states. Therefore, although the edge states do not appear in our calculations, they have very limited influence on CPGE in the cases we considered in this paper.

At critical thickness $\mathrm{L}_\mathrm{c1}$, there is no substantial change in CPGE when the HgTe QW transits from BI to TI. The reason is that although the first conduction and valence subbands exchange their components, they do not make an impact on the transition probability between them as predicted by Fermi's golden rule. Interestingly, we find the CPGE could be greatly enhanced at a certain range of the spectrum when HgTe QW enters the heavily inverted regime, i.e., $\mathrm{L}_\mathrm{w}>\mathrm{L}_\mathrm{c2}$.
This is because in the inverted phase regime, H1 and H2 become the first conduction and valence subband, while H2 has a distorted M-like energy dispersion due to the strong coupling with H1 and E1. For example, in a 9-nm HgTe QW [see Fig. \ref{fig3}(d)], we can see H2 unusually bends upwards in the range of  $0<|\mathbf{k}|<0.4$ nm$^{-1}$. This distorted dispersion gives rise to a remarkable increase of joint density of states at the corresponding energy spectrum (40-100 meV). As a result, the optical absorption and CPGE current are evidently enhanced (about two-four times larger in the 9-nm HgTe QW). Note that this enhancement agrees very well with the CPGE signal rise in the wavelength ranges of 12-15 $\mu$m as reported in Ref. \onlinecite{Wittman_HgTeCPGE_2010}. This feature could be an experimental evidence that HgTe QWs are in the heavily inverted regime and coexist with the topological edge states \cite{Bernevig_QSHE_2006}. In addition, this effect could be advantageous in the design of a high-efficiency ellipticity detector of radiation in infrared and terahertz regions \cite{Ganichev_THzPLDetect_2008,Danilov_THZdetecor_2009}.

\subsection{Influence of bulk inversion asymmetry\label{sec:BIAeffect}}

Hg$_{1-x}$Cd$_{x}$Te has a zinc-blende structure, which lacks a center of inversion and give rise to the BIA spin splitting. In HgTe QWs, because BIA is considered to be much smaller than SIA, the BIA effect is less explored. However, introducing BIA terms in the Hamiltonian would lead to a qualitatively different symmetry of system. This could be reflected on the CPGE currents.

In this section, we discuss the influence of BIA on CPGE by adding the BIA term $\hat{H}_{BIA}$, i.e., setting $\hat{H}_{A}=\hat{H}_{SIA}+\hat{H}_{BIA}$. The BIA term is known as the Dresselhaus $k^3$ term for the parabolic conduction band model. For the eight-band model, there are two kinds of BIA terms, i.e., the Kane's off-diagonal terms with parameters $B_{8v}^{\pm},B_{7v}$ \cite{Kane1966,Winkler2003}, and $\mathbf{k}$-linear terms in the $\Gamma_8$ block of the Hamiltonian with parameter $C_{k}$ \cite{Cardona_LinearK_1986,Cardona_LinearK_1988}, respectively. These BIA terms can be derived by the theory of invariants \cite{Trebin_Invariants_1979,Winkler2003}. The form of $\hat{H}_{BIA}$ and BIA parameters are presented in Appendix \ref{Apdx:BIA}.

\begin{figure} [tpb]
\includegraphics[width=0.85\columnwidth]{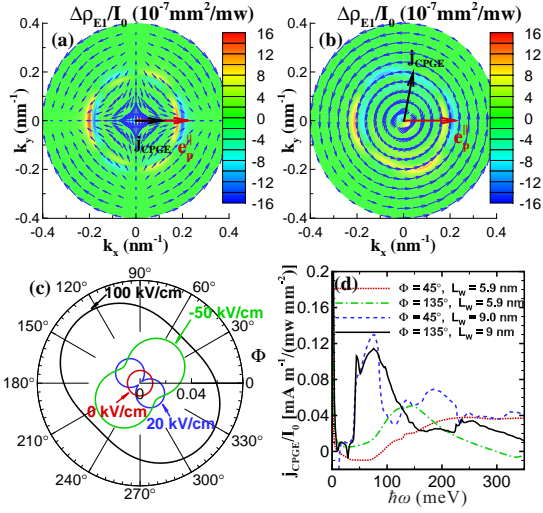}% Here is how to import EPS art
\caption{\label{fig4} (Color online)
(a) Calculated $\bm{\mathfrak{B}}_{E1}(\mathbf{k})$ (minor blue arrows) and $\Delta \rho_{E1}(\mathbf{k})$ (contour color) for a 5.9-nm HgTe QW with BIA only. The incident light is right-hand circularly polarized, with $\hbar \omega =$ 129 meV, $\Phi=$ 0$^\circ$, and $\Theta _{0}=$ 30$^\circ$. (b) The same as (a) but the QW has both BIA and SIA ($\mathcal{F}=$ 80 kV/cm). (c) The magnitude of CPGE current [in unit of mA m$^{-1}\times I_0$/(mW mm$^{-2}$)] as a function of the incident azimuth angle $\Phi$, for a 5.9-nm QW with BIA and different SIA electric field $\mathcal{F}=$ 0, 20, 100 and -50 kV/cm. The photon energy is $\hbar \omega =$ 160 meV. (d) The CPGE current spectrums for $\Phi=$ 45$^\circ$ ($\mathbf{e}_p^{\parallel} \parallel [110]$) and 135$^\circ$ ($\mathbf{e}_p^{\parallel} \parallel [\bar{1}10]$), in a 5.9- and 9-nm HgTe QW ($\mathcal{F} =$ 60 kV/cm), respectively.}
\end{figure}

For HgTe QWs with BIA only, i.e., let $\hat{H}_{A}=\hat{H}_{BIA}$, the symmetry of the system belongs to the $D_{2d}$ point group, as reflected in the effective magnetic field of the E1 subband, i.e., $\mathfrak{B}_{E1}(\mathbf{k})$ in Fig. \ref{fig4}(a).
Because $\mathfrak{B}_{E1}(\mathbf{k})$ of BIA is parallel (antiparallel) to $\mathbf{k}$ at the $[100]$ ($[\bar{1}00]$) direction, if $\Phi =$ 0$^\circ$, i.e., $\mathbf{e}_p^{\parallel} \parallel$ $[100]$, the $\sigma^{+}$ light would excite asymmetrical distribution of $\Delta \rho_{mn}(\mathbf{k})$ along the $[100]$-$[\bar{1}00]$ direction [see the diagonal element of $\Delta \rho_{mn}(\mathbf{k})$ for the E1 subband, i.e., $\Delta \rho_{E1}(\mathbf{k})$ in Fig. \ref{fig4}(a)]. As a result, a net charge current along the $[100]$ direction will be produced, which corresponds to a nonzero pseudotensor component $\gamma _{xx}$. Symmetry analysis shows $\gamma _{xx}=-\gamma_{yy}$ are the only nonzero components of $\gamma _{\lambda \mu}$ for QWs with BIA only.

For HgTe QWs with both BIA and SIA, the symmetry is reduced to the $C_{2v}$ point group, as displayed by $\mathfrak{B}_{E1}(\mathbf{k})$ in Fig. \ref{fig4}(b). The $\sigma^{+}$ light with $\mathbf{e}_p^{\parallel} \parallel [100]$ could excite CPGE currents with both $[100]$ and $[010]$ components, because $\mathfrak{B}_{E1}(\mathbf{k})$ can be viewed as the superposition of RSOI and DSOI magnetic effective fields. In this case, there are two types of independent components of $\gamma _{\lambda \mu}$, which are $\gamma_{xx}=-\gamma _{yy}$ and $\gamma _{yx}=-\gamma _{xy}$, respectively.

Due to the interference of RSOI and DSOI, the CPGE currents exhibit anisotropic behavior about the azimuthal angle $\Phi$ of light [see in Fig. \ref{fig4}(c)]. We can see the minimum and maximum of CPGE currents appear at $\Phi =$ 45$^{\circ}$ or 135$^{\circ}$, and the degree of anisotropy can be effectively tuned by changing the electric field $\mathcal{F}$. One can achieve the strongest anisotropy of CPGE currents at certain conditions when $|\gamma_{xx}| \approx |\gamma_{yx}|$ [e.g., see the blue line for $\mathcal{F}=$ 20 kV/cm in Fig. \ref{fig4}(c)]. Under these conditions, the RSOI happens to cancel DSOI at $\mathbf{k} \parallel [110]$ (or $[\bar{1}10]$), so that there could be plenty of interesting phenomena such as the suppression of weak antilocalization \cite{Knap_WAL_1996}, the disappearance of SdH oscillation beating \cite{Averkiev_SDH_2005}, the very long spin relaxation time for spins oriented $[110]$ or $[\bar{1}10]$ \cite{Averkiev_SRT_1999}, and the persistent spin helix \cite{Bernevig_PSH_2006,Koralek_PSH_2009}. The anisotropy of CPGE about the azimuthal angle of incident light could offer another means to find the conditions when RSOI cancels DSOI.

In Fig. \ref{fig4}(d), we plot the calculated CPGE current spectra for incident light with $\mathbf{e}_p^{\parallel} \parallel [110]$ and $[\bar{1}10]$ in 5.9-nm and 9-nm HgTe QWs, respectively. In general, the spectra of CPGE currents $\mathbf{e}_p^{\parallel} \parallel [110]$ and $\mathbf{e}_p^{\parallel} \parallel [\bar{1}10]$ are not equal, but there are crossing points in spectra where CPGE currents are equal (which means CPGE could be isotropic at certain photon energies). Comparing the CPGE spectra of 5-nm and 9-nm HgTe QWs, we find in 5-nm QW the difference of spectra for $\mathbf{e}_p^{\parallel} \parallel [110]$ and $\mathbf{e}_p^{\parallel} \parallel [\bar{1}10]$ is larger. This is because the influence of BIA is more prominent in narrower QWs, implying the anisotropy behavior of CPGE can be more easily observed in narrow QWs.

\begin{figure} [tpb]
\includegraphics[width=0.85\columnwidth]{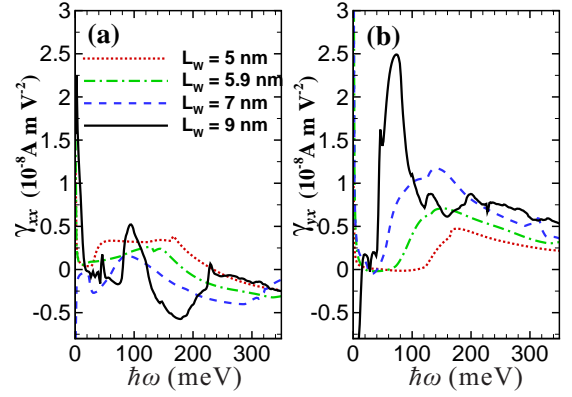}
\caption{\label{fig5} (Color online) (a) and (b) The spectrums of $\gamma_{xx}$ and $\gamma_{yx}$, respectively, for 5-, 5.9-, 7-, 9-nm HgTe QWs with both BIA and SIA ($\mathcal{F}=$ 80 kV/cm).}
\end{figure}

Substituting the components of $\mathbf{E}$ in Eq. (\ref{eq:J6}) by Eq. (\ref{LEvertor}), we can obtain a simple phenomenological expression for the photocurrents
 \begin{eqnarray} \label{Jsimp}
j_{x} &=&E_{0}^{2}t_{s}t_{p}\sin \Theta P_{circ}( \gamma _{xx}\cos \Phi
-\gamma _{yx}\sin \Phi ) , \notag \\
j_{y} &=&-E_{0}^{2}t_{s}t_{p}\sin \Theta P_{circ}( \gamma _{xx}\sin
\Phi -\gamma _{yx}\cos \Phi ) .
\end{eqnarray}
If we choose the configuration $x\parallel[100]$ and $y\parallel[100]$ [as in Fig. \ref{fig1}(a)], by using Eq. (\ref{Jsimp}) we can demonstrate the ratio of CPGE currents satisfying
\begin{equation}
\frac{j_{y}(\mathbf{e}_p^{\parallel} \parallel x ) }{j_{x}(\mathbf{e}_p^{\parallel} \parallel x)}=\frac{j_{x}(\mathbf{e}_p^{\parallel} \parallel y) }{j_{y}(\mathbf{e}_p^{\parallel} \parallel y)}=\frac{\gamma
_{yx}}{\gamma _{xx}}.
\end{equation}
If we choose another configuration, i.e., $x'\parallel[110]$ and $y'\parallel[\bar{1}10]$, we can find
$j_{x'}(\mathbf{e}_p^{\parallel} \parallel x')=j_{y'}(\mathbf{e}_p^{\parallel} \parallel y')=0$, and
\begin{equation}
\frac{j_{y'}(\mathbf{e}_p^{\parallel} \parallel x')}{j_{x'}(\mathbf{e}_p^{\parallel} \parallel y') }=\frac{\gamma _{xx}-\gamma
_{yx}}{\gamma _{xx}+\gamma_{yx}}.
\end{equation}
Giglberger \textit{et al}. have shown that $\gamma_{yx}/\gamma_{xx}$ obtained by CPGE is very close to the RD ratio measured by spin-galvanic effect \cite{Giglberger_RDration_2007}. Since then CPGE has been widely used in determining  the RD ratio \cite{Ganichev_RDratio_2004,Giglberger_RDration_2007, CMYin_GaNCPGE_2010, CYH_RDQWrs_2011}.
However, we should mention that $\gamma_{yx}/\gamma_{xx}$ may not always
equal to the RD ratio, because $\gamma _{\lambda \mu}$ have very complex dependence on the photon energy.
Our calculation results [Fig. \ref{fig5}] show  CPGE currents are not only dependent on the spin splitting of conduction and valence bands, but also on the joint density of states and the number of subbands involved in the optical transition, which makes the spectra of $\gamma _{\lambda \mu}$ very complicated. For example, we can see there are independent peaks and sign reversions in the $\gamma_{xx}$ and $\gamma_{yx}$ spectra of 9-nm HgTe QW, which lead to the value of $\gamma_{yx}/\gamma_{xx}$ varying in a wide range. Therefore, for materials with band-structure abnormalities like 9-nm HgTe QWs, $\gamma_{yx}/\gamma_{xx}$ may not applicable in determining the RD ratio. To analyze the experimental results of CPGE and extracting unusual band structure information, theory based on a multi-band model is necessary.

\section{Pure spin currents induced by linearly polarized light \label{sec:PSC}}

A pure spin current is usually defined by a spin flow without net charge current.
In non-centrosymmetric semiconductors, PSCs can be generated by illuminating a single-color linearly polarized light on the sample \cite{Bhat_PSC_2005,Ivchenko_PSC_2005,ZhaoHui_PSC_2005}.
 As linearly polarized light can be regarded as the coherent superposition of two circularly polarized lights with opposite helicities and equal strengths, it can drive equal numbers of spin-up and spin-down carriers traveling in the opposite directions as by CPGE. In this situation, the net charge currents are cancelled but the pure spin flows are formed. This method provides an optical means to excite PSCs into semiconductors.

In this section, by setting $\varphi =0$ and $\Theta_{0}=0$ in Eq. (\ref{LEvertor}), one can get a linearly polarized light at normal incidence.
The linear polarization direction of light can be changed by $\Phi$, and $\Phi=$ 0$^\circ$ gives a light linearly polarized along the $[110]$ direction.
Then the PSCs driven by linearly polarized light in HgTe QWs can be investigated within the same theoretical framework of density-matrix formalism.
By utilizing the second-order steady-state density matrix, the excited PSCs can be calculated by
\begin{equation}\label{PSC}
j_{\alpha }^{\beta }=\sum_{\mathbf{k},m,n}\rho_{mn}^{(2)\infty}(\mathbf{k})
\langle n,\mathbf{k}\vert \hat{j}_{\alpha }^{\beta
}\vert m,\mathbf{k}\rangle.
\end{equation}%
Here $j_{\alpha }^{\beta}$ stands for the spin current moving along the $\alpha$ direction and spins orienting in the $\beta$ direction, and $\hat{j}_{\alpha }^{\beta}$ is the spin current operator. In this paper, we adopt the standard definition of spin current operator \cite{Rashba_EquiliSC_2003}, i.e., $\hat{j}_{\alpha }^{\beta}\equiv \frac{\hbar }{4}( \hat{v}_{\alpha}\hat{\Sigma}_{\beta}+\hat{\Sigma}_{\beta}\hat{v}_{\alpha})$, where $\hat{v}_{\alpha}$ is the $\alpha$ component of the velocity operator defined in Eq. (\ref{velocityop}), and $\hat{\Sigma}_{\beta}$ is the $\beta$ component of eight-band spin matrices \cite{Junli_spinstates_2010}. Note that the definition of spin current operator is still a controversial issue \cite{ShiJunRen_DefinationSC_2006}. By the standard definition, Rashba first found out there are nonzero equilibrium spin currents existing in systems with SOIs, and they are not directly corresponding to the transport of spins which could lead to spin accumulation \cite{Rashba_EquiliSC_2003}. However, there are papers suggesting the standard spin current operator makes physical sense and does not need to be modified, because the equilibrium spin currents can be viewed as the persistent spin flows similar to the persistent Meissner currents \cite{Sonin_EquiliSC_2007, SunQF_PSCring_2008, Tokatly_EquiliSC_2008}. Here, we stick to the standard definition of spin currents to be consistent with the previous works about the PSCs driven by linearly polarized light \cite{Bhat_PSC_2005,Ivchenko_PSC_2005,Ivchenko_PSC_2008,ZhouBin_CPGE_2007}. But one should remember that the PSCs obtained by Eq. (\ref{PSC}) are similar to the equilibrium spin currents, except they are excited by linearly polarized light [note that in Eq. (\ref{PSC}) we already use $\rho_{mn}^{(2)\infty}(\mathbf{k})$ instead of $\rho_{mn}(\mathbf{k})$ to exclude the equilibrium spin currents]. Like the equilibrium spin currents, the optically excited PSCs may not directly result in the spin accumulation at the edges of the sample. But the exited PSCs are measurable values, and can be detected by second-order nonlinear optical effects \cite{WangJing_NOESC_2010,ZhaoHui_NOESC_2010}, as well as the change of  mechanical torques near edges of sample \cite{Sonin_EquiliSC_2007} and the electric field in a ring device induced by equilibrium spin currents \cite{SunQF_PSCring_2008}.

The phenomenological expression of photogalvanic spin currents is written as
\begin{equation}
j_{\alpha }^{\beta }=\sum_{\gamma \delta }\mu _{\alpha \beta \gamma \delta
}E_{\gamma }E_{\delta }^{\ast }.
\end{equation}
We only consider the in-plane spin currents induced by normal incidence of light, therefore $\alpha, \beta, \gamma, \delta \in \{x, y\}$.
$\mu _{\alpha \beta \gamma \delta}$ is a fourth-rank tensor. For the linearly polarized light, $E_{\gamma }E_{\delta }^{\ast }\equiv E_{\gamma
}E_{\delta }$ is real, which restricts $\mu _{\alpha \beta \gamma \delta}$ to be also real and
symmetric with respect to the interchange of the last two indices, i.e., $\mu _{\alpha \beta \gamma \delta}=\mu _{\alpha \beta \delta \gamma }$.
Using Eq. (\ref{rho2}), we can derive the microscopic expression for $\mu _{\alpha \beta \gamma \delta }$ as
\begin{eqnarray}\label{muPSC}
\mu _{\alpha \beta \gamma \delta } = \frac{e^{2}}{\omega ^{2}\hbar ^{2}} \sum_{\mathbf{k}} \sum_{mnq}( f_{m}-f_{q})\times[j_{\alpha ,nm}^{\beta}v_{mq}^{\gamma }v_{qn}^{\delta }\mathcal{L}_{m,n,q}(\omega)+j_{\alpha ,mn}^{\beta }v_{nq}^{\gamma }v_{qm}^{\delta }\mathcal{L}^{\ast}_{m,n,q}(\omega) \nonumber \\
+j_{\alpha ,nm}^{\beta }v_{qn}^{\gamma }v_{mq}^{\delta }\mathcal{L}_{m,n,q}(-\omega)+j_{\alpha ,mn}^{\beta }v_{qm}^{\gamma }v_{nq}^{\delta}\mathcal{L}^{\ast }_{m,n,q}(-\omega)],
\end{eqnarray}
where $j_{\alpha ,mn}^{\beta }\equiv \langle m,\mathbf{k}\vert
\hat{j}_{\alpha }^{\beta }\vert n,\mathbf{k}\rangle$ and the definition of $\mathcal{L}_{m,n,q}(\omega)$ is the same as in Eq. (\ref{chi}).

\begin{figure} [tpb]
\includegraphics[width=0.85\columnwidth]{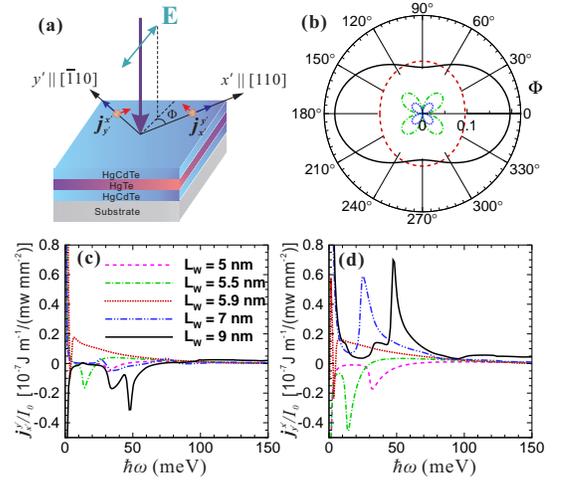}
\caption{\label{fig6}(Color online) (a) Schematic of configuration for the generation of PSCs in a HgTe QW by normally irradiating the linearly polarized light. (b) The four PSCs [in unit of 10$^{-7}$J m$^{-1}\times$I$_0$/(mw mm$^{-2}$)]: $j_{x'}^{y'}$ (red dashed line), $j_{y'}^{x'}$ (black solid line), $j_{x'}^{x'}$ (green dash-dotted line) and $j_{y'}^{y'}$ (blue dotted line) as a function of the linear polarization direction of light, in a 9-nm HgTe QW. The incident photon energy is $\hbar \omega =$ 45 meV. (c) and (d) The spectra of $j_{x'}^{y'}$ and $j_{y'}^{x'}$ induced by [110] linearly polarized light, respectively, for HgTe QWs with different well widths. In all above panels, we have assumed the HgTe QWs have both BIA and SIA ($\mathcal{F}=$ 80 kV/cm).}
\end{figure}

Symmetry analysis shows that there could be six independent components of $\mu_{\alpha \beta \gamma \delta}$ for a general HgTe QW.
They are
\begin{eqnarray*}
\mu _{1} &=&\mu _{xxxx}=-\mu _{yyyy}, \\
\mu _{2} &=&\mu _{yyxx}=-\mu _{xxyy}, \\
\mu _{3} &=&\mu _{xyxx}=-\mu _{yxyy}, \\
\mu _{4} &=&\mu _{yxxx}=-\mu _{xyyy}, \\
\mu _{5} &=&\mu _{xxxy}=-\mu _{yyxy}=\mu _{xxyx}=-\mu _{yyyx}, \\
\mu _{6} &=&\mu _{xyxy}=\mu _{xyyx}=-\mu _{yxxy}=-\mu _{yxyx}.
\end{eqnarray*}
If there is only SIA in the system, we find $\mu _{1},\mu _{2},\mu _{6}=0$ but
$\mu _{3},\mu _{4},\mu _{5}\neq 0$, which means the $[100]$ linearly polarized light could excite two nonzero PSCs, i.e., $j_{x}^{y}$ and $j_{y}^{x}$, respectively.
Otherwise, if only BIA exists, we find $\mu _{1},\mu _{2},\mu
_{6}\neq 0,$ but $\mu _{3},\mu _{4},\mu _{5}=0$. $j_{x}^{x}$ and $j_{y}^{y}$ would be the nonzero PSCs driven by $[100]$ linearly polarized light.
If both SIA and BIA are present, all the components $\mu_{1}, ..., \mu_{6}$ are nonzero.
So $[100]$-linearly-polarized light would give rise to four nonzero PSCs, which are $j_{x}^{x}$, $j_{x}^{y}$, $j_{y}^{x}$ and $j_{y}^{y}$, respectively.

Alternatively, we can choose another configuration with $x'\parallel[110]$ and $y'\parallel[\bar{1}10]$ for the description of PSCs [as sketched in Fig. \ref{fig6}(a)]. The relationship between $j_{\alpha'}^{\beta'}$ ($\alpha',\beta'\in \{x',y'\}$) and $j_{\alpha}^{\beta}$ ($\alpha,\beta \in \{x,y\}$) can be found by coordinates rotation. Let $\mathbb{R}(\vartheta)$ be the in-plane rotation matrix of a rotation angle $\vartheta$
\begin{equation}
\mathbb{R}(\vartheta)=\left(
\begin{array}{cc}
\cos \vartheta  & \sin \vartheta  \\
-\sin \vartheta  & \cos \vartheta
\end{array}%
\right).
\end{equation}
Then the relation is found to be
\begin{equation}
j_{\alpha'}^{\beta'}=\sum_{\alpha \beta }\mathbb{R}_{\alpha' \alpha }(\frac{\pi }{4})\mathbb{R}_{\beta'\beta }(\frac{\pi }{4}
)j_{\alpha }^{\beta}.
 \end{equation}
In Fig. \ref{fig6}(b) we plot the four in-plane PSCs, i.e., $j_{x'}^{y'}$, $j_{y'}^{x'}$, $j_{x'}^{x'}$, and $j_{y'}^{y'}$, in 9 nm HgTe QW as a function of the light's linear polarized direction $\Phi$.
We can see $j_{x'}^{x'}$ and $j_{y'}^{y'}$ are zero when the incident light is linearly polarized along $[110]$ ($\Phi=$ 0$^\circ$, 180$^\circ$) or $[\bar{1}10]$ ($\Phi=$ 90$^\circ$, 270$^\circ$), but they have finite values along other directions.
For QWs with both BIA and SIA, the four PSCs are not equivalent to each other, since the symmetry of QW is reduced to the $C_{2v}$ point group.

If we restrict the incident light linearly polarized along $[110]$ (or $[\bar{1}10]$) direction, we can find $j_{x'}^{y'}$ and $j_{y'}^{x'}$ are the only nonzero PSCs regardless of the interplay of SIA and BIA. The spectra of $j_{x'}^{y'}$ and $j_{y'}^{x'}$ induced by $[110]$ linearly polarized light for HgTe QWs with different well widths are displayed in Fig. \ref{fig6}(c) and \ref{fig6}(d). We can see the spectra of $j_{x'}^{y'}$ and $j_{y'}^{x'}$ show sharp peaks near the optical absorption edges. At photon energies above the absorption edges, the PSCs decrease quickly. The reason is that the band mixing at large $\mathbf{k}$ greatly suppresses the expectation values of spin currents. In addition, we find the band mixing effect is more prominent in $j_{x'}^{y'}$ than in $j_{y'}^{x'}$ for $[110]$ linearly polarized light. Therefore the magnitudes of $j_{x'}^{y'}$ are generally smaller than that of $j_{y'}^{x'}$. Interestingly, we find $j_{y'}^{x'}$ change signs if we increase $\mathrm{L}_\mathrm{w}$ across the critical thickness of TI phase transition, i.e., $\mathrm{L}_\mathrm{c1}$. Or in other words, $j_{y'}^{x'}$ have different signs for HgTe QWs with BI phase and TI phase. This is because both the contributions of E1 and H1 states to $j_{y'}^{x'}$ change signs when the order of E1 and H1 is reversed. This feature implies $j_{y'}^{x'}$ might be utilized to characterize the quantum phase of HgTe QWs. In contrast, $j_{x'}^{y'}$ is not so sensitive to quantum phase transition because of the strong band mixing effect smeared out the change of the subband's character.

\section{Conclusion \label{sec:conclusion}}

In summary, we presented a theoretical method for the calculation of circular photogalvanic charge currents and linearly photogalvanic pure spin currents based on the eight-band $\mathbf{k\cdot p}$ model and density-matrix formalism.
This method could take account of the complex band structure details and different type of inversion asymmetries, and is used to investigate the CPGE currents and PSCs and their microscopic origins in HgTe QWs with different quantum phases. Our calculations show CPGE could be remarkably enhanced at a certain range of the energy spectrum due to the distorted band structures of heavily inverted HgTe QWs. The interference of RSOI and DSOI could lead to the CPGE currents anisotropically dependent on the azimuthal angle of incident light. For QWs with abnormal band structures, $\gamma _{\lambda \mu }$ have very complicated dependency on the spin splittings and band dispersions, so $\gamma_{yx}/\gamma_{xx}$ does not simply equal to the RD ratio. For $[110]$-linearly-polarized light at normal incidence, the light could drive two nonzero PSCs, i.e., $j_{x'}^{y'}$ and $j_{y'}^{x'}$ ($x'\parallel[110]$ and $y'\parallel[\bar{1}10]$), respectively. We find $j_{y'}^{x'}$ are different in signs for HgTe QWs with BI phase and TI phase. These findings are helpful for understanding the experimental results and designing novel HgTe-based infrared and terahertz optoelectronic devices.

\begin{acknowledgments}
This work was supported by the National Natural
Science Foundation of China (Grant No. 11104232), the Fundamental Research Funds for the Central Universities (Grant No. 20720160019), and the Natural Science Foundation of Fujian Province of China (Grant No. 2016J05163).
W. Y. was supported by the NSFC (Grants No. 11274036 and No. 11322542), the MOST (Grants No. 2014CB848700), and the NSFC program for "Scientific Research Center" (Grant No. U1530401).
J.-T. L. was supported by the National Natural Science Foundation of China (Grant No. 11364033).
We would like to thank Prof. Kai Chang for inspiring suggestions.
\end{acknowledgments}

\appendix
\section{Eight-Band Hamiltonian\label{Apdx:Hamiltonian}}

The band structure of narrow gap QWs can be well described within the framework of Burt's envelope function formalism together with Kane's eight-band
$\mathbf{k\cdot p}$ Hamiltonian \cite{Novik_HgMnTeBand_2005}. The exact form of the Hamiltonian is dependent on the choice of basis set.
In this work, the eight-band basis set is chosen as
\begin{eqnarray}\label{eqn:basis}
\phi _{1}& = &\left \vert \frac{1}{2},\frac{1}{2}\right \rangle =\left \vert
S\uparrow \right \rangle ,  \nonumber \\
\phi _{2}& = &\left \vert \frac{1}{2},-\frac{1}{2}\right \rangle =\left \vert
S\downarrow \right \rangle , \nonumber \\
\phi _{3}& = &\left \vert \frac{3}{2},\frac{3}{2}\right \rangle =\frac{1}{\sqrt{2%
}}\left \vert (X+iY)\uparrow \right \rangle , \nonumber \\
\phi _{4}& = &\left \vert \frac{3}{2},\frac{1}{2}\right \rangle =\frac{i}{\sqrt{6%
}}\left \vert (X+iY)\downarrow -2Z\uparrow \right \rangle ,\nonumber \\
\phi _{5}& = &\left \vert \frac{3}{2},-\frac{1}{2}\right \rangle =\frac{1}{\sqrt{%
6}}\left \vert (X-iY)\uparrow +2Z\downarrow \right \rangle , \nonumber \\
\phi _{6}& = &\left \vert \frac{3}{2},-\frac{3}{2}\right \rangle =\frac{i}{\sqrt{%
2}}\left \vert X-iY\downarrow \right \rangle , \nonumber \\
\phi _{7}& = &\left \vert \frac{1}{2},\frac{1}{2}\right \rangle =\frac{1}{\sqrt{3%
}}\left \vert (X+iY)\downarrow +Z\uparrow \right \rangle , \nonumber \\
\phi _{8}& = &\left \vert \frac{1}{2},-\frac{1}{2}\right \rangle =-\frac{i}{%
\sqrt{3}}\left \vert (X-iY)\uparrow -Z\downarrow \right \rangle ,
\end{eqnarray}

\begin{table*}[ptb]
\centering
\begin{ruledtabular}%
\caption{The band parameters used in our calculations. These
parameters are taken from Ref.~\onlinecite{Novik_HgMnTeBand_2005} and~\onlinecite{LBBook}. }%
\begin{tabular}[c]{ccccccccccc}
& $E_{v}$ (eV) & $E_{g}$ (eV) &$E_{p}=2m_0P_0^2/\hbar^2$ (eV)& $\Delta$ (eV) & F & $\gamma_{1}$ & $\gamma_{2}$ & $\gamma_{3}$ & $\kappa$ &$n_r$\\
\hline
HgTe \cite{Novik_HgMnTeBand_2005} & 0 & -0.303 & 18.8 &1.08 & 0 & 4.1 & 0.5 & 0.3 & -0.4 & 3.28 \\
Hg$_{0.3}$Cd$_{0.7}$Te \cite{LBBook} &-0.399 & 1.006 & 18.8 & 1.0 & -0.8 & 3.3 & 0.1 & 0.9 & -0.8 & 3.28 \\
\end{tabular}
\label{tab:para}%
\end{ruledtabular}
\end{table*}

In the presentation of this basis set, the eight-band Hamiltonian $\hat{H}_{K}$ is
\begin{widetext}
\begin{equation}\small
\begin{bmatrix}% \small
\mathcal{A} & 0 & i\sqrt{3}\mathcal{V}^{\dagger } & \sqrt{2}\mathcal{U} & i%
\mathcal{V} & 0 & i\mathcal{U} & \sqrt{2}\mathcal{V} \\
0 & \mathcal{A} & 0 & -\mathcal{V}^{\dagger } & i\sqrt{2}\mathcal{U} & -%
\sqrt{3}\mathcal{V} & i\sqrt{2}\mathcal{V}^{\dagger } & -\mathcal{U} \\
-i\sqrt{3}\mathcal{V} & 0 & -(\mathcal{P}+\mathcal{Q}) & \mathcal{L}-i\sqrt{3%
}\mathcal{N} & \mathcal{M} & 0 & \frac{i}{\sqrt{2}}\mathcal{L}+\sqrt{\frac{3%
}{2}}\mathcal{N} & -i\sqrt{2}\mathcal{M} \\
\sqrt{2}\mathcal{U} & -\mathcal{V} & \mathcal{L}^{\dag }-i\sqrt{3}\mathcal{N}%
^{\dag } & -(\mathcal{P}-\mathcal{Q}) & -2i\mathcal{N} & \mathcal{M} & i%
\sqrt{2}\mathcal{Q} & i\sqrt{\frac{3}{2}}\mathcal{L}-\sqrt{\frac{1}{2}}%
\mathcal{N} \\
-i\mathcal{V}^{\dag } & -i\sqrt{2}\mathcal{U} & \mathcal{M}^{\dag } & 2i%
\mathcal{N}^{\dag } & -(\mathcal{P}-\mathcal{Q}) & -\mathcal{L}-i\sqrt{3}%
\mathcal{N} & -i\sqrt{\frac{3}{2}}\mathcal{L}^{\dag }+\sqrt{\frac{1}{2}}%
\mathcal{N}^{\dag } & i\sqrt{2}\mathcal{Q} \\
0 & -\sqrt{3}\mathcal{V}^{\dag } & 0 & \mathcal{M}^{\dag } & -\mathcal{L}%
^{\dag }+i\sqrt{3}\mathcal{N}^{\dag } & -(\mathcal{P}+\mathcal{Q}) & -i\sqrt{%
2}\mathcal{M}^{\dag } & -\frac{i}{\sqrt{2}}\mathcal{L}^{\dag }-\sqrt{\frac{3%
}{2}}\mathcal{N}^{\dag } \\
-i\mathcal{U} & -i\sqrt{2}\mathcal{V} & -\frac{i}{\sqrt{2}}\mathcal{L}^{\dag
}+\sqrt{\frac{3}{2}}\mathcal{N}^{\dag } & -i\sqrt{2}\mathcal{Q} & i\sqrt{%
\frac{3}{2}}\mathcal{L}-\sqrt{\frac{1}{2}}\mathcal{N}^{\dag } & i\sqrt{2}%
\mathcal{M} & -\mathcal{P}-\Delta  & 2i\mathcal{N} \\
\sqrt{2}\mathcal{V}^{\dag } & -\mathcal{U} & i\sqrt{2}\mathcal{M}^{\dag } &
-i\sqrt{\frac{3}{2}}\mathcal{L}^{\dag }-\sqrt{\frac{1}{2}}\mathcal{N}^{\dag }
& -i\sqrt{2}\mathcal{Q} & \frac{i}{\sqrt{2}}\mathcal{L}-\sqrt{\frac{3}{2}}%
\mathcal{N} & -2i\mathcal{N}^{\dag } & -\mathcal{P}-\Delta
\end{bmatrix}
\label{eqn:Hamiltonian}
\end{equation}
\end{widetext}
where
\begin{eqnarray}\label{eq:Helements}
\mathcal{A} & = &E_{v}+E_{g}+\frac{\hbar ^{2}}{2m_{0}}[(2F+1)k^{2}+\hat{k}_{z}(2F+1)\hat{k}_{z}] ,\nonumber \\
\mathcal{P} & = &-E_{v}+\frac{\hbar ^{2}}{2m_{0}}( \gamma_{1}k^{2}+\hat{k}_{z}\gamma _{1}\hat{k}_{z}) , \nonumber\\
\mathcal{Q} & = &\frac{\hbar ^{2}}{2m_{0}}(\gamma _{2}k^{2}-2\hat{k}_{z}\gamma_{2}\hat{k}_{z}), \nonumber \\
\mathcal{L} & = &i\frac{2\sqrt{3}\hbar ^{2}}{m_{0}}k_{-}\{\gamma _{3},\hat{k}_{z}\}, \nonumber \\
\mathcal{M} & = &-\frac{\sqrt{3}\hbar ^{2}}{2m_{0}}[\gamma _{2}(k_{x}^{2}-k_{y}^{2}) -i2\gamma _{3}k_{x}k_{y}], \nonumber\\
\mathcal{N} & = &\frac{\hbar ^{2}}{2m_{0}}k_{-}[\hat{k}_{z},\kappa] ,\nonumber \\
\mathcal{U} & = &\frac{1}{\sqrt{3}}P_{0}\hat{k}_{z},\nonumber \\
\mathcal{V} & = &\frac{1}{\sqrt{6}}P_{0}k_{-}.
\end{eqnarray}

For the (001)-oriented HgTe QW, $\hat{k}_z$ should be replaced by $\hat{k}_z \rightarrow -i\partial
/\partial _{z}$ as a result of quantum confinement. $\{ \hat{A},\hat{B}\} = (\hat{A}\hat{B}+\hat{B}\hat{A})/2$
and $[\hat{A},\hat{B}] = \hat{A}\hat{B}-\hat{B}\hat{A}$ denote the anticommutator and usual commutator for operators $\hat{A}$ and $\hat{B}$.
$\mathbf{k}\equiv(k_{x},k_{y})$ is the in-plane wave vector, $k^2\equiv k_{x}^2+k_{y}^2$, and $k_{\pm }\equiv k_{x}\pm ik_{y}$.
The band-structure parameters, including
$E_v$, $E_g$, $P_0$, $F$, $\gamma_1$, $\gamma_2$, $\gamma_3$, and $\kappa$,
are dependent on the materials of each layers. These parameters for HgTe and Hg$_{0.3}$Cd$_{0.7}$Te are listed in Table \ref{tab:para}.
In the Hamiltonian of heterostructures, the parameters can be assumed as the step functions along growth direction $z$.
In our calculation, $\hat{H}_{K}$ is taken as the Hamiltonian of symmetric HgTe QWs,
which means the step functions of parameters have mirror reflection symmetry.
As a consequence, $\hat{H}_{K}$ holds the spatial inversion symmetry, so it does not give rise to the spin spitting of the band structure.

For a general HgTe QW with Hamiltonian $\hat{H}_{0}$, the eigenenergy and the eigenstate of electron with wave vector $\mathbf{k}$ can be obtained by solving the time-independent
Schr\"{o}dinger equation
\begin{equation}
\hat{H}_{0}|m,\mathbf{k}\rangle =\varepsilon _{m}(\mathbf{k})|m,\mathbf{k}\rangle .
\label{eq:sch}
\end{equation}
Here $m$ is the subband index, $\varepsilon_m(\mathbf{k})$ is the eigenenergy, and
$|m,\mathbf{k}\rangle$ is the eigenstate. $|m,\mathbf{k}\rangle$ is a vector with eight components of envelope functions $|m,\mathbf{k}\rangle =\exp(i\mathbf{k}\cdot\mathbf{r})[\varphi _{1}^{m}(z),\varphi _{2}^{m}(z),...,\varphi_{8}^{m}(z)]^{T}$. Equation (\ref{eq:sch}) is equivalent to a system of coupled differential
equations, which can be solved by the plane wave expansion method, i.e., one can
expand each envelope function $\varphi_{n}^{m}(z)$ as a series of plane waves
\begin{equation}
\varphi_{n}^{m}(z)=\frac{1}{\sqrt{L}}\sum_{j=-N}^{N}c_{nj}^{m}\exp
(ik_{j}z),  \label{eq:pwex}
\end{equation}
where $k_{j}=2j\pi /L$ and $L$ is the total length of the structure,
and $N$ is the cut-off plane wave number. By moderately choosing $N$ ($N=40$ is used in this work), one can avoid the spurious solutions \cite{Wyang_SpuriousSolution_2005} as well as getting results with required accuracy.
Substituting Eq.(\ref{eq:pwex}) into Eq.(\ref{eq:sch}), the coupled differential
equations are then converted to the standard eigenvalue problem which can be numerically solved by matrix diagonalization.

\section{Bulk Inversion Asymmetry terms\label{Apdx:BIA}}
For the eight-band model, there are two kinds of terms which could give birth to BIA, i.e., terms weighted by $B_{8v}^{\pm},B_{7v}$ \cite{Kane1966,Winkler2003} and $C_{k}$ \cite{Cardona_LinearK_1986,Cardona_LinearK_1988}, respectively. The terms with $B_{8v}^{\pm}$ ($B_{7v}$) come from the indirect coupling between $\Gamma_6$ and $\Gamma_8$ ($\Gamma_7$) bands mediated by the remote bands. These terms are quadratic in $\mathbf{k}$, and appear in the off-diagonal blocks of Hamiltonian. The terms with  $C_{k}$ are linear in $\mathbf{k}$ and present in the $\Gamma_8$ block of the Hamiltonian (called $\Gamma_8$ band $\mathbf{k}$-linear terms). They mainly come from the second-order perturbation terms combining the matrix elements of $\mathbf{k\cdot p}$ and the spin-orbit operator $\hat{H}_{SO}$. The values of $B_{8v}^{\pm},B_{7v}$ can be evaluated from the 14-band Hamiltonian of the extended Kane model \cite{Winkler2003}, and the values of $C_{k}$ have been studied by Cardona \textit{et al}. \cite{Cardona_LinearK_1986}. Here, for Hg$_{1-x}$Cd$_{x}$Te, we neglect the difference between $B_{8v}^{+}$ and $B_{7v}$, and assume $B_{+} \simeq (B_{8v}^{+}+B_{7v})/2$, $B_{-} = B_{8v}^{-}$. These BIA parameters are presented in Table \ref{tab:BIApara}.
\begin{table}[ptb]
\centering
\begin{ruledtabular}
\caption{The BIA parameters used in our calculations. These
parameters are obtained from in Ref.~\onlinecite{Winkler2003,Cardona_LinearK_1986,Winkler_HgTeBIA_2012}.}
\begin{tabular}
[c]{cccc} & $B_{+}$ (eV$\cdot${\AA}$^2$)& $B_{-}$ (eV$\cdot${\AA}$^2$)& $C_{k}$ (eV$\cdot${\AA})\\ \hline
HgTe & -20.0 & 1.0 & -0.0746\\
CdTe & -21.44 & -0.635 & -0.0234\\
\end{tabular}
\label{tab:BIApara}
\footnotetext[1] {The BIA parameters for Hg$_{x}$Cd$_{1-x}$Te are assumed to be the linear interpolation of the parameters of HgTe and CdTe.}
\end{ruledtabular}
\end{table}

In the representation of the eight-band basis [Eq. (\ref{eqn:basis})], the form of $\hat{H}_{BIA}$ is
\begin{widetext}
\begin{equation}
\hat{H}_{BIA}=%
\begin{pmatrix}
0 & 0 & i\sqrt{3}\mathcal{W}_{2}^{\dag } & -\sqrt{2}(\mathcal{W}_{1}+%
\mathcal{T}_{1}) & -i\mathcal{W}_{2} & \mathcal{T}_{2} & -i\mathcal{W}_{1} &
-\sqrt{2}\mathcal{W}_{2} \\
0 & 0 & -i\mathcal{T}_{2} & -\mathcal{W}_{2}^{\dag } & -i\sqrt{2}(\mathcal{W}%
_{1}-\mathcal{T}_{1}) & \sqrt{3}\mathcal{W}_{2} & i\sqrt{2}\mathcal{W}%
_{2}^{\dag } & \mathcal{W}_{1} \\
-i\sqrt{3}\mathcal{W}_{2} & i\mathcal{T}_{2}^{\dag } & 0 & \mathcal{C}_{1} & 2\mathcal{C}_{2} &
\sqrt{3}\mathcal{C}_{1}^{\dag } & \frac{i}{\sqrt{2}}\mathcal{C}_{1} & i\sqrt{2}\mathcal{C}_{2} \\
-\sqrt{2}(\mathcal{W}_{1}^{\dag }+\mathcal{T}_{1}^{\dag }) & -\mathcal{W}_{2}
& \mathcal{C}_{1}^{\dag } & 0 & -\sqrt{3}\mathcal{C}_{1} & -2\mathcal{C}_{2} & 0 & i\sqrt{\frac{3}{2}}\mathcal{C}_{1}
\\
i\mathcal{W}_{2}^{\dag } & i\sqrt{2}(\mathcal{W}_{1}^{\dag }-\mathcal{T}%
_{1}^{\dag }) & 2\mathcal{C}_{2}^{\dag } & -\sqrt{3}\mathcal{C}_{1}^{\dag } & 0 & \mathcal{C}_{1} & i\sqrt{%
\frac{3}{2}}\mathcal{C}_{1}^{\dag } & 0 \\
\mathcal{T}_{2}^{\dag } & \sqrt{3}\mathcal{W}_{2}^{\dag } & \sqrt{3}\mathcal{C}_{1} &
-2\mathcal{C}_{2}^{\dag } & \mathcal{C}_{1}^{\dag } & 0 & -i\sqrt{2}\mathcal{C}_{2} & \frac{i}{\sqrt{2}}%
\mathcal{C}_{1}^{\dag } \\
i\mathcal{W}_{1}^{\dag } & -i\sqrt{2}\mathcal{W}_{2} & -\frac{i}{\sqrt{2}}%
\mathcal{C}_{1}^{\dag } & 0 & -i\sqrt{\frac{3}{2}}\mathcal{C}_{1} & i\sqrt{2}\mathcal{C}_{2}^{\dag } & 0 &
0 \\
-\sqrt{2}\mathcal{W}_{2}^{\dag } & \mathcal{W}_{1}^{\dag } & -i\sqrt{2}%
\mathcal{C}_{2}^{\dag } & -i\sqrt{\frac{3}{2}}\mathcal{C}_{1}^{\dag } & 0 & -\frac{i}{\sqrt{2}}%
\mathcal{C}_{1} & 0 & 0%
\end{pmatrix}%
\end{equation},
\end{widetext}
where
\begin{eqnarray}
\mathcal{W}_{1}&=&-\frac{i}{\sqrt{3}}B_{+}k_{x}k_{y}, \nonumber \\
\mathcal{W}_{2}&=&-\frac{1}{\sqrt{6}}k_{+}\{B_{+},\hat{k}_{z}\}, \nonumber \\
\mathcal{T}_{1}&=&-\frac{1}{2\sqrt{3}}B_{-}( k_{x}^{2}-k_{y}^{2}),\nonumber \\
\mathcal{T}_{2}&=&-\frac{1}{3\sqrt{2}}[B_{-}(k_{x}^{2}+k_{y}^{2})-\hat{k}_{z}B_{-}\hat{k}_{z}],\nonumber \\
\mathcal{C}_{1}&=&-\frac{1}{2}iC_{k}k_{+},\nonumber \\
\mathcal{C}_{2}&=&-\frac{1}{2}\{C_{k},\hat{k}_{z}\}.
\end{eqnarray}

In the elements of $\hat{H}_{BIA}$, $\hat{k}_{z}$ is also replaced by $\hat{k}_z \rightarrow -i\partial
/\partial _{z}$ as in Eq. (\ref{eq:Helements}). $\hat{H}_{BIA}$ could give a small modification of $\hat{H}_{K}$, and can be included in $\hat{H}_{0}$ and together solved by Eq. (\ref{eq:sch}).

\section{Effective Magnetic Field of Spin-Orbit Interactions\label{Apdx:EMF}}
The spin-orbit coupling originates from the relativistic transformation of electric field and magnetic field. In the reference frame of a moving electron, a static electric field is transformed into a magnetic field depending on the velocity
(or the wave vector $\mathbf{k}$) of the electron. The electron spin could couple to this transformed effective magnetic field via the magnetic dipole interaction.
In this sense, the effects of SOI can be understood directly by analogy with a $\mathbf{k}$-dependent effective magnetic field $\bm{\mathfrak{B}}(\mathbf{k})$. Similar to the Zeeman effect of a real magnetic field, the effective magnetic field of SOI could also split the energy band into two branches, with the spin orientation of the upper (lower) branch parallel (antiparallel) to the direction of $\bm{\mathfrak{B}}(\mathbf{k})$.
In different bands, the electron may feel SOI with different strengths. Therefore we should label the effective magnetic field of SOI felt by the electron in the $n$th subband with wave vector $\mathbf{k}$ as $\bm{\mathfrak{B}}_n(\mathbf{k})$.
The SOI-induced spin splitting and electron spin orientations will be closely dependent on the magnitudes and directions of $\bm{\mathfrak{B}}_n(\mathbf{k})$ respectively. Due to $\bm{\mathfrak{B}}_n(\mathbf{k})$ changes with $\mathbf{k}$, the spin splitting and spin orientations change as well, producing the spin texture of the $n$th subband in $\mathbf{k}$ space.

The effective magnetic field can be defined by attributing the spin splitting of SOI to the Zeeman effect of $\bm{\mathfrak{B}}_n(\mathbf{k})$.
In the eight-band basis, the Zeeman term is written as \cite{Winkler2003}
\begin{equation}
\hat{H}_{z}=\mu _{B}\mathbf{B\cdot\bm{\hat{\mathcal{J}}} },
\end{equation}
where $\mu _{B}$ is the Bohr magneton, $\mathbf{B}$ is the external magnetic field, and $\bm{\hat{\mathcal{J}}}=(\hat{\mathcal{J}}_x,\hat{\mathcal{J}}_y,\hat{\mathcal{J}}_z)$ is the vector of
eight-band angular momentum matrices \cite{ZMtermsnote}. The form of $\bm{\hat{\mathcal{J}}}$ can be found in Ref. \onlinecite{Winkler2003}.
Letting $\bm{\mathfrak{B}}_n(\mathbf{k})=\mathbf{B}$, we can obtain
\begin{equation}\label{eq:DeltaE}
\Delta \varepsilon_{n}(\mathbf{k})=\varepsilon_{n+}(\mathbf{k})-\varepsilon_{n-}(\mathbf{k})\approx
\mu _{B} \bm{\mathfrak{B}}_n(\mathbf{k})\mathbf{\cdot}\bm{\mathcal{J}}_n(\mathbf{k)}.
\end{equation}
$\varepsilon_{n\pm}(\mathbf{k})$ are the energies for the upper (lower) branch of the $n$th subband and $\Delta \varepsilon_{n}(\mathbf{k})$ is the spin splitting.
$\bm{\mathcal{J}}_{n}(\mathbf{k)}\equiv \langle n+,\mathbf{k}|
\bm{\hat{\mathcal{J}}}|n+,\mathbf{k}\rangle -\langle n-,\mathbf{k}|\bm{\hat{\mathcal{J}}}|n-,\mathbf{k}%
\rangle$, and $\vert n\pm,\mathbf{k}\rangle$ are the eigenstates for the upper (lower) branch of the $n$th subband.
Using Eq. (\ref{eq:DeltaE}), we can get the expression for the effective magnetic field of the $n$th subband
\begin{equation}\label{eq:EMF}
\bm{\mathfrak{B}}_n(\mathbf{k})= \frac{\Delta \varepsilon_{n}(\mathbf{k})\bm{\mathcal{J}}_{n}(\mathbf{k)}}{\mu _{B}\vert\bm{\mathcal{J}}_{n}(\mathbf{k)}\vert^2 }.
\end{equation}

Equation (\ref{eq:EMF}) indicates that the magnitudes and directions of $\bm{\mathfrak{B}}_n(\mathbf{k})$ could represent the spin spitting and spin orientations (or spin texture) of $n$th subband, respectively. For the analysis of SOI-induced phenomena, it is very helpful to visualize the $\mathbf{k}$ space distributions of effective magnetic fields.

\bibliography{CPGE_in_HgTe_QWs}

\begin{thebibliography}{68}
\bibitem [{ZMt()}]{ZMtermsnote}%
  \BibitemOpen
  \href@noop {} {\ }\bibinfo {note} {For materials in a real magnetic field,
  the eight-band spin matrices in Zeeman terms should contain the effective g
  factor parameter $\kappa$. Howerver, we find the difference with and without
  $\kappa$ are small. Besides, we are acctually interested in the relative
  magnitude of effective magnetic field, so the terms with $\kappa$ is
  negelcted here.}\BibitemShut {Stop}%
\end{thebibliography}

\end{document}